\documentclass{PoS}
\usepackage{amsmath}
\usepackage{cite}

\title{SM vacuum stability and the Weyl consistency conditions: Counting to three}

\ShortTitle{SM vacuum stability and the Weyl consistency conditions: Counting to three}

\author{\speaker{Jens Krog}%
        \\
       CP3-Origins, University of Southern Denmark\\
       E-mail: \email{krog@cp3-origins.net}}

\abstract{We demonstrate how a new perturbative ordering may result from the structure of the Weyl anomaly. Respecting the abelian nature of the Weyl anomaly at the lowest order enforces the use of beta functions calculated to a different loop order for different types of couplings. These consistency conditions are found to be satisfied by the renormalization group equations of the standard model, and we perform an analysis of the vacuum stability of the Higgs potential respecting the consistency conditions and compare to the previous results. Hints toward unknown structure in the standard model renormalization group equations are found, although the vacuum stability results are in agreement with previous estimates.}

\FullConference{Proceedings of the Corfu Summer Institute 2014 \\
		 3-21 September 2014\\
		 Corfu, Greece}

\begin{document}

\section{Introduction}

The renormalization group (RG) equations are becoming an increasingly important tool for theorists to understand and develop new theories. Extrapolating between different energy scales is essential in theories, where the dynamics play a crucial part. In many theories the infrared (IR) phenomena are directly coupled to the running of the couplings as in Technicolor-like theories, theories with a Coleman-Weinberg mechanism, as well as QCD. In other cases, the ultraviolet (UV) limit is of the essence, as in theories of asymptotic safety or the recent investigations of the stability of the standard model Higgs potential. In any case, the RG analysis can be an essential part of the investigation, and as such the precision of this analysis is vital for the quality of the results.

Obtaining high precision in an RG analysis is usually associated with using calculations incorporating a large number of loops, thus reducing the error associated to the truncation of the infinite series of diagrams contributing to the RG equations. When several RG equations are in play, the standard procedure is to calculate as many loops as possible for each of the equations. As the beta functions for gauge and Yukawa couplings are usually simpler to calculate than the ones of any quartic scalar coupling, the "loop order" of the beta functions involved in the analysis is sometimes not the same. An analysis where all beta functions are calculated to X loops, is often referred to as being a "X-loop" calculation, or equivalently a $\text{N}^{X-1}$LO calculation.

This method of "ordering" calculations originates in the notion that quantum loops contribute with powers of $\hbar$, and by keeping all RG equations at the same loop order, a consistent expansion in $\hbar$ is performed. As the inclusion of further loops in reality adds terms of successively higher power of the coupling constants, one could however argue that loop ordering should be done according to some measure of the relative size of the couplings, so it is not clear that a specific ordering is correct. The further argument, which will be essential to the content of the following sections, is that the standard ordering is not consistently balancing the interplay of the different couplings. At one loop, for example, the evolution of the scalar self coupling may be affected by both the gauge and Yukawa couplings, while the gauge coupling only receives corrections from itself. In the following, we will see that a natural ordering may be found, which balances the influence of these couplings, by examining the structure of the Weyl anomaly - an anomaly resulting directly from the breaking of scale symmetry, which is exactly what drives the running of the couplings in the first place.

\section{The Weyl Anomaly}
We will take our starting point in the soon to be defined Weyl anomaly, following roughly the work of Osborn \cite{Osborn:1991gm}.
Starting from a conformal field theory, we add marginal operators $\mathcal {O}^i$ to the theory;
\begin{equation}
\mathcal{L}=\mathcal{L}_{CFT}+g_i\mathcal{O}^i \, .
\end{equation}
These operators have dimensionless coupling constants $g_i$, and the theory is classically scale invariant.

As a tool for the analysis ahead, we will add a nontrivial background metric $\gamma^{\, \mu\nu}$ and promote the coupling constants to being space-time dependent:
\begin{align}
\gamma^{\, \mu\nu} &\rightarrow \gamma^{\, \mu\nu}(x) \nonumber \\
g \, &\rightarrow \, g(x). \nonumber
\end{align}
We now perform a \textit{local} scale transformation signified by the parameter $\sigma(x)$, which transforms the couplings $g_i$ and the metric as
\begin{align}
\gamma^{\, \mu\nu} &\rightarrow e^{2\sigma (x)}\gamma^{\, \mu\nu}  \\
g_i(\mu) \, &\rightarrow \, g_i(e^{-\sigma(x)}\mu). 
\end{align}
As the theory is only classically scale invariant, this transformation will result in the variation of the generating functional at the quantum level, which is defined by
\begin{equation}
W= \text{log} \left[\int \mathcal{D}\Phi \, e^{i\int \textrm{d}^4x \, \mathcal{L}} \right]
\end{equation}
This variation is exactly the Weyl anomaly, and it is given by
\begin{equation}
\Delta_\sigma W \equiv \int \textrm{d}^4x \, \sigma(x)
		\left( 2 \gamma_{\mu\nu} \frac{\delta W}{\delta \gamma_{\mu\nu}}
			- \beta_i \frac{\delta W}{\delta g_i} \right)
		= \sigma \left( a E(\gamma) 
			+ \chi^{ij} \partial_\mu g_i \partial_\nu g_j G^{\mu\nu} \right)
			+ \partial_\mu \sigma \omega^i \, \partial_\nu g_i G^{\mu\nu} 
			+ \ldots
\label{Wanom}
\end{equation}
Here, $G^{\mu\nu}$ is the Einstein tensor, $E$ is the Euler tensor, $\beta_i$ is the beta functions for the coupling $g_i$, and $a$, $\omega^i$, and $\chi^{ij}$ are functions of the couplings $g_i$. Naturally, reverting to constant values for $g_i$ and a trivial metric, would trivialize this equation somewhat, but what we are interested in is mainly the structure to be found in the functions $a$, and $\chi^{ij}$ and their relation to the beta functions.

An important feature of the group of Weyl transformations is that it is abelian, such the order of two successive transformations is irrelevant
\begin{equation}
\Delta_\sigma \Delta_\tau W = \Delta_\tau \Delta_\sigma W.
\end{equation}
When applying this condition to the anomaly \eqref{Wanom}, a set of consistency conditions arise, among which
\begin{equation}
\frac{\partial \tilde{a}}{\partial g_i}=\left(-\chi^{ij}+\frac{\partial \omega^i}{\partial g_j}-\frac{\partial \omega^j}{\partial g_i} \right)\beta_j \equiv -\chi^{ij}\beta_j,
\label{dadg}
\end{equation}
where we have defined $\tilde{a}=a-\omega^i\beta_i$.
Regarding $\tilde{\chi}$ as a sort of metric for the coupling constants, we now define the "contravariant" 
\begin{equation}
\beta^i\equiv \tilde{\chi}^{ij}\beta_i,
\end{equation}
and we can rewrite \eqref{dadg} as
\begin{equation}
\frac{\partial \tilde{a}}{\partial g_i} = -\beta^i,
\end{equation}
applying a second partial derivative, we may then conclude:
\begin{equation}
\frac{\partial^2 \tilde{a}}{\partial g_i \partial g_j} = -\frac{\partial \beta^i}{\partial g_j} \quad \Leftrightarrow \quad \frac{\partial \beta^i}{\partial g_j} = \frac{\partial \beta^j}{\partial g_i}.
\label{betacc}
\end{equation}
Evidently, the abelian nature of the Weyl group is reflected directly in a set of conditions linking the "contravariant" beta functions to each other, which we will henceforth refer to as the Weyl consistency conditions. Specifically, these conditions will be between cross terms in the beta functions, that is terms in $\beta^i$ containing $g_j$ and vice versa. This set of relations will also be present in the limit where the couplings are restored to constants and the metric goes towards the trivial one, and is fundamentally present in any field theory which contains only marginal couplings\footnote{This statement can be generalized to theories where dimensional couplings are present, but where the renormalization procedure is a mass independent one, as in $\overline{\text{MS}}$.}.
In the following, we will use the standard model (in an $\overline{\text{MS}}$ scheme, where the Higgs mass is ignored) as an example, and show that these relations are explicitly fulfilled to lowest order.

\section{Perturbative counting and the Standard Model}
To see the consequences of the relations described above, we turn to the standard model (SM). We will restrict our attention to the case where only the three gauge couplings ($g_1,g_2,g_3$), the top Yukawa coupling $(y_t)$, and the scalar self coupling of the Higgs ($\lambda$) are nonzero. This allows for full use of the framework described while only working with the couplings mostly relevant to the RG analysis due to their large values in the SM.

For simplicity we redefine the couplings, such that the resulting SM couplings we work with are
\begin{equation}
\alpha_1=\frac{g_1^2}{(4\pi)^2} \, ,\quad \alpha_2=\frac{g_2^2}{(4\pi)^2} \, ,\quad \alpha_3=\frac{g_3^2}{(4\pi)^2} \, ,\quad \alpha_t=\frac{y_t^2}{(4\pi)^2} \, ,\quad \alpha_\lambda=\frac{\lambda}{(4\pi)^2} \, .
\end{equation}
The RG analysis will be carried out for RG scales $\mu \gg M_t$, where $M_t$ is the top mass, such that all particles can be considered massless, thus making the theory classically scale invariant and  enabling the use of the framework described in the previous section.

The beta functions of these couplings are known to three loop order\cite{Mihaila:2012fm,Bednyakov:2012rb,Bednyakov:2012en,Chetyrkin:2012rz,Chetyrkin:2013wya}. To check whether or not they satisfy the Weyl consistency conditions, \eqref{betacc}, we need to calculate $\tilde{\chi}_{ij}$, which will be done only to the lowest nontrivial order.
At the lowest order, the functions $\omega^i$ are exact one forms, such that $\frac{\partial \omega^i}{\partial g_j} -\frac{\partial \omega^j}{\partial g_i}=0$, and so $\tilde{\chi}=\chi$. A calculation of $\chi$ at the lowest order is performed using vacuum diagrams, and we find
\begin{equation}
\chi = \text{diag} \left(\frac{1}{\alpha_1^2},\frac{1}{\alpha_2^2},\frac{3}{\alpha_3^2},\frac{1}{\alpha_t},4 \right ) \, .
\label{chidiag}
\end{equation}
The important feature of this expression is that different types of couplings enter the equations with different powers, the gauge couplings to power minus two, the top coupling to power minus one and the quartic enters at power zero. Inserting this result in \eqref{betacc}, where we define the beta functions as $\beta_i=\mu^2 \frac{\partial^2 \alpha_i}{\partial \mu^2}$, we find the set of relations:
\begin{eqnarray}
	2 \frac{\partial}{\partial\alpha_t} \beta_\lambda
		& = & \frac{\partial}{\partial\alpha_\lambda} \left( \frac{\beta_t}{\alpha_t} \right) + \mathcal{O}\left( \alpha_i^2 \right)
		\label{eq:consistencycondition:lambda:t} \\
	4 \frac{\partial}{\partial\alpha_1} \beta_\lambda
		& = & \frac{\partial}{\partial\alpha_\lambda} \left( \frac{\beta_1}{\alpha_1^2} \right) + \mathcal{O}\left( \alpha_i^2 \right)
		\label{eq:consistencycondition:lambda:1} \\
	\frac{4}{3} \frac{\partial}{\partial\alpha_2} \beta_\lambda
		& = & \frac{\partial}{\partial\alpha_\lambda} \left( \frac{\beta_2}{\alpha_2^2} \right) + \mathcal{O}\left( \alpha_i^2 \right)
		\label{eq:consistencycondition:lambda:2} \\
	2 \frac{\partial}{\partial\alpha_1} \left( \frac{\beta_t}{\alpha_t} \right)
		& = & \frac{\partial}{\partial\alpha_t} \left( \frac{\beta_1}{\alpha_1^2} \right) + \mathcal{O}\left( \alpha_i^2 \right)
		\label{eq:consistencycondition:t:1} \\
	\frac{2}{3} \frac{\partial}{\partial\alpha_2} \left( \frac{\beta_t}{\alpha_t} \right)
		& = & \frac{\partial}{\partial\alpha_t} \left( \frac{\beta_2}{\alpha_2^2} \right) + \mathcal{O}\left( \alpha_i^2 \right)
		\label{eq:consistencycondition:t:2} \\
	\frac{1}{4} \frac{\partial}{\partial\alpha_3} \left( \frac{\beta_t}{\alpha_t} \right)
		& = & \frac{\partial}{\partial\alpha_t} \left( \frac{\beta_3}{\alpha_3^2} \right) + \mathcal{O}\left( \alpha_i^2 \right)
		\label{eq:consistencycondition:t:3}\\
	\frac{1}{3} \frac{\partial}{\partial\alpha_2} \left( \frac{\beta_1}{\alpha_1^2} \right)
		& = & \frac{\partial}{\partial\alpha_1} \left( \frac{\beta_2}{\alpha_2^2} \right) + \mathcal{O}\left( \alpha_i^2 \right)
		\label{eq:consistencycondition:1:2} \\
	\frac{1}{8} \frac{\partial}{\partial\alpha_3} \left( \frac{\beta_1}{\alpha_1^2} \right)
		& = & \frac{\partial}{\partial\alpha_1} \left( \frac{\beta_3}{\alpha_3^2} \right) + \mathcal{O}\left( \alpha_i^2 \right)
		\label{eq:consistencycondition:1:3} \\
	\frac{3}{8} \frac{\partial}{\partial\alpha_3} \left( \frac{\beta_2}{\alpha_2^2} \right)
		& = & \frac{\partial}{\partial\alpha_2} \left( \frac{\beta_3}{\alpha_3^2} \right) + \mathcal{O}\left( \alpha_i^2 \right)
		\label{eq:consistencycondition:2:3}
\end{eqnarray}

A careful look at the beta functions in question reveals that these relations are all respected, thus supplying a consistency check for the original calculations \cite{Antipin:2013sga}. What should be noted, as evident from the structure of \eqref{chidiag}, is that the relations are not between terms in the beta function of the same loop order, but rather connects terms of different loop orders for different types of couplings.

Explicitly, one can see from \eqref{eq:consistencycondition:lambda:t}, \eqref{eq:consistencycondition:lambda:1}, and \eqref{eq:consistencycondition:lambda:2}, that the consistency conditions \eqref{betacc} relate the mixed terms in the one loop beta function for the quartic coupling to the two loop terms in the Yukawa beta function and to the three loop gauge beta functions. Had we only calculated the gauge beta functions to one or two loops, and the Yukawa beta function to one loop, this relation would thus not have been satisfied, even though only the one loop beta function was used for the quartic coupling.
Similarly, one can see from \eqref{eq:consistencycondition:t:1}, \eqref{eq:consistencycondition:t:2}, and \eqref{eq:consistencycondition:t:3}, that the consistency conditions relate the two loop terms in the top beta function to the three loop terms in the gauge beta functions. 

In order to ensure that the Weyl anomaly, measuring the departure from scale invariance, is correctly of abelian nature, we thus see that the for the lowest nontrivial order calculation one needs the quartic beta functions calculated to one loop, the Yukawa beta functions to two loops, and the gauge beta functions to three loops ("3-2-1" ordering). In this setup then, a conventional calculation to one loop in all couplings is not a complete lowest order calculation at all! In addition, in order to specify a consistent NLO counting, one needs to calculate corrections to the lowest order "metric", and it is not certain that a consistent order expansion can be found. In any case, it can be said that under the "3-2-1" ordering, the requirement of an abelian Weyl anomaly is satisfied to lowest order, while a conventional ordering leads to a departure from Weyl consistency. 

\section{Vacuum stability reviewed}

As we have described a new lowest order ordering which respects the Weyl consistency conditions, we should investigate how big the effect from departing from this ordering is. If departing from Weyl consistency means large changes in the predictions for the theory in question, it becomes vital to understand which result should be trusted. For the standard model, the UV behavior of the theory has been studied in great detail, and specifically the (in)stability of the Higgs vacuum has been questioned.

The conventional lowest order calculations, using only one loop beta functions, has been performed long ago. On the other hand, a calculation using three loop beta functions for all of the mentioned couplings has recently been performed \cite{EliasMiro:2011aa}.

In the analysis of the Higgs vacuum stability, the most important parameter is the Higgs quartic coupling, $\lambda$. We calculate the RG flow as the renormalization scale changes from the electroweak scale to the Planck scale. If the Higgs quartic coupling becomes negative at any scale in between, this can be seen as a sign of deviation from total stability\footnote{For a more precise calculation, one must consider the flow of an effective $\lambda_{eff}$, defined via the effective potential: $V_{eff}(\phi)=\lambda_{eft}(\phi)/4\phi^4$, where we have exchanged the renormalization scale for the Higgs field value $\phi$.}.

\begin{figure}[b]
\begin{center}
\includegraphics[width=0.43\textwidth]{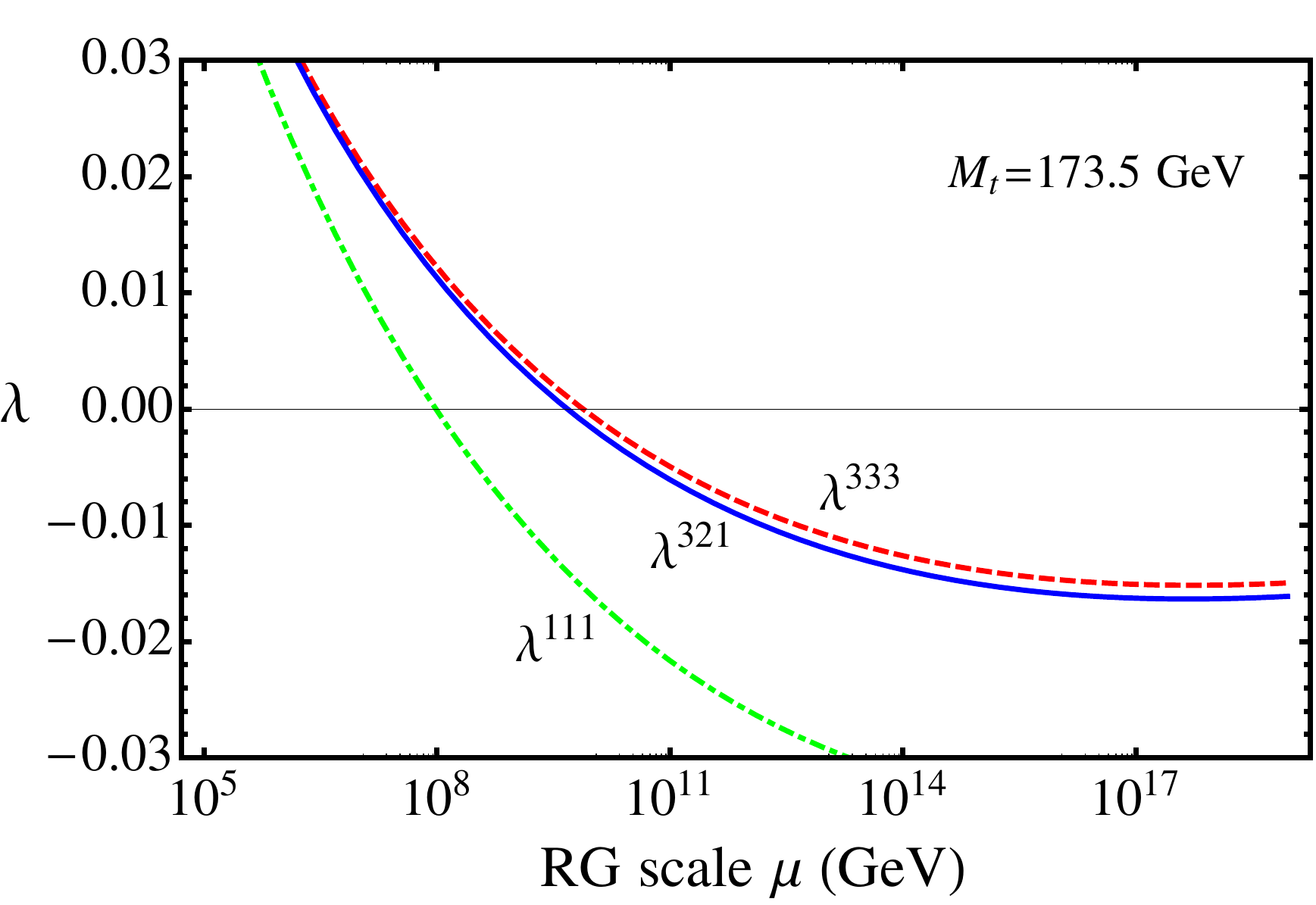}
\hspace{.5cm}
\includegraphics[width=0.43\textwidth]{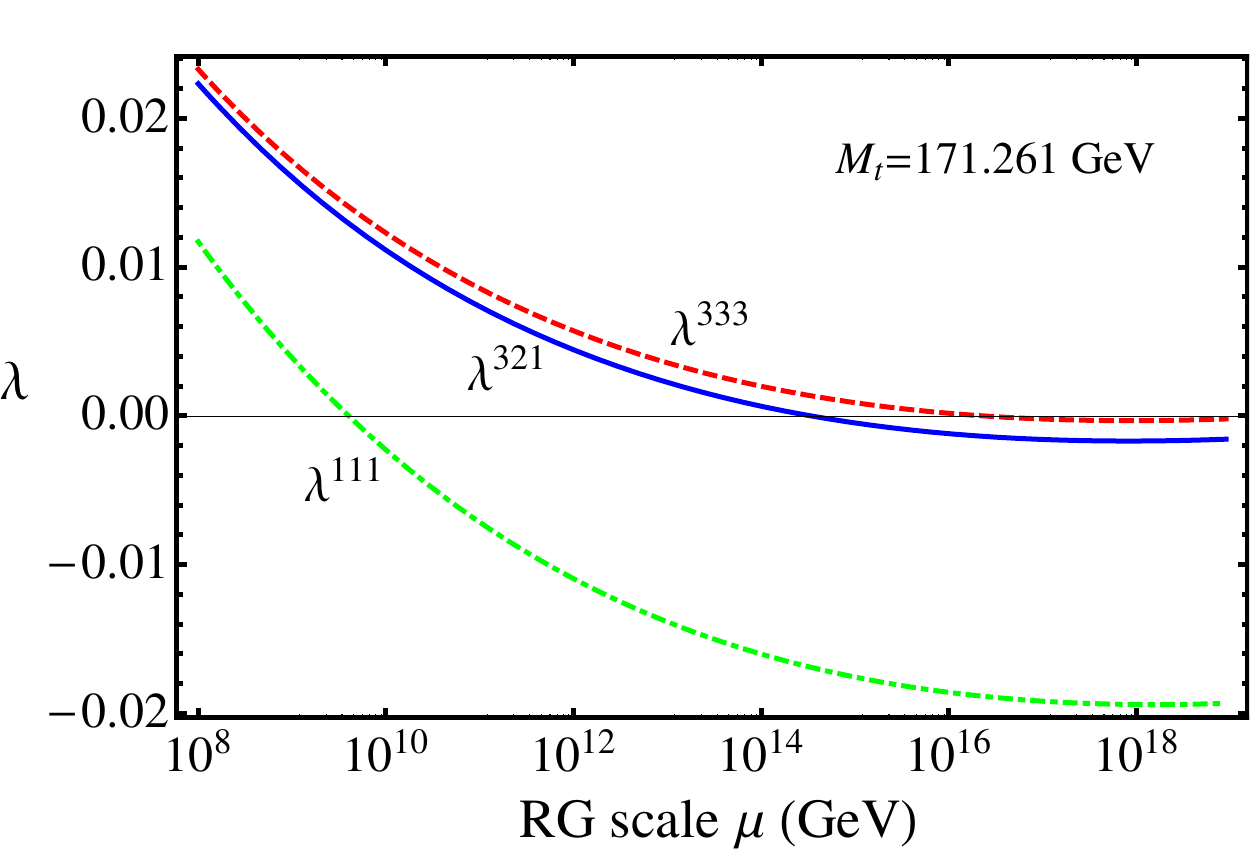}
\label{quartrg}
\caption{The RG evolution of the quartic Higgs coupling in different orderings, where the Higgs mass is set at its central value $M_H =125.7$ GeV and the top mass is set to $M_t=173.5 (171.261)$ GeV to the left(right).}
\end{center}
\end{figure}

We show the evolution of the Higgs quartic coupling $\lambda$ in three cases: 
\begin{description} 
\item{\bf{(1-1-1)}} The conventional leading order result, where all beta functions are calculated to one loop.
\item{\bf{(3-2-1)}} The lowest order calculation respecting the Weyl consistency conditions, using the three loop gauge beta functions, the two loop top Yukawa beta function, and the one loop quartic beta function
\item{\bf{(3-3-3)}} The conventional "NNLO" calculation using three loop beta functions for all couplings.
\end{description}
The resulting evolutions are shown in Fig.~\ref{quartrg} for the central values of the Higgs and top mass, and for the value of the top mass which results in $\lambda=0$ at the Planck scale.
Evidently, there is a large difference when upgrading from the conventional lowest order (1-1-1) to the Weyl consistent lowest order (3-2-1). Note that this is despite the fact, that the beta function for the quartic coupling is identical in the two orderings. It is harder to draw conclusions when comparing to the full three loop analysis (3-3-3). Clearly the two methods yield approximately the same result for the running of the quartic coupling, even though the quartic beta function for the latter has a considerably larger amount of terms. This feature may be due to large cancellations in the higher order terms for the quartic coupling, which has already been pointed out by the authors of \cite{Chetyrkin:2013wya}. This tells us that the influence of the extra loops in the gauge and top Yukawa beta functions is greater than the higher order terms of the quartic itself, even for the quartic evolution.

As we are not in a position to say if any ordering can be Weyl consistent beyond the leading order, it is not clear how to make a consistent comparison between the NLO or NNLO methods, or to estimate how badly the conditions \eqref{betacc} are violated in the 3-3-3 ordering. It is, however, worthwhile to note that it seems that the gauge and Yukawa couplings have far greater influence on the running of the quartic coupling than naively expected. Perhaps one should therefore hope to obtain greater precision by calculating further corrections to the gauge and Yukawa beta functions, and not the quartic itself.

\begin{figure}[b]
\begin{center}
\includegraphics[width=0.7\textwidth]{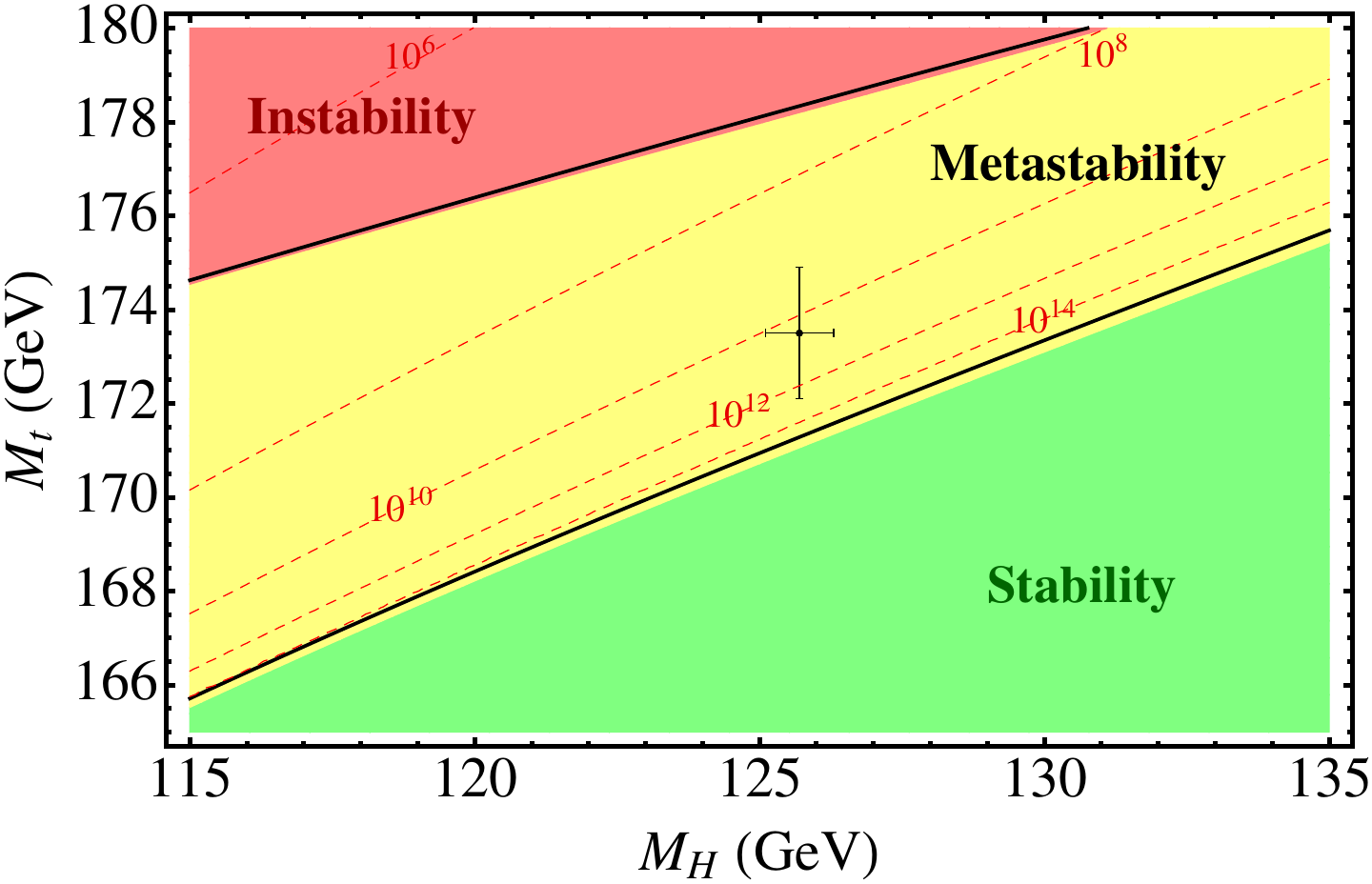}
\label{money}
\caption{The phase diagram of the standard model, where regions of instability, metastability, and stability of the Higgs potential are labeled, as a function of the Higgs and top masses. The solid lines represent the separation of the phases in the 3-3-3 ordering, while the coloring represents the Weyl consistent 3-2-1 ordering.}
\end{center}
\end{figure}

The stability analysis for the Higgs potential is usually carried out as a scan over the top and Higgs mass, and we include our analysis, where we compare the 3-3-3 and 3-2-1 orderings in Fig.~\ref{money}.
The calculation involves calculation of the effective potential as well as the tunneling rate in case a second minimum exists at higher field values. Metastability indicates that the average life time of the electroweak minimum is longer than the current age of the Universe. As it is seen in the figure, the two methods yield approximately the same results. The discrepancy might become important, however, once greater precision is obtained for the values of the top and Higgs mass.

\section{Conclusion}
We have demonstrated how a new perturbative ordering arises from the loss of scale invariance due to quantum effects by examining the structure of the Weyl anomaly. Specifically, to obtain consistency with the abelian nature of the Weyl anomaly to leading order in general quantum field theories, we have shown that the beta function of the quartic coupling must be calculated to one loop, while the beta functions for the Yukawa and gauge coupling must be calculated to two and three loops respectively. The relation between the beta functions have been found to exist within the SM, and we have compared the RG evolution of the Higgs potential within different orderings. We see that the higher order modifications to the gauge and Yukawa runnings have a much larger impact than the higher order terms in the Higgs quartic itself. Although the conventional NNLO result is not very different from the leading order Weyl consistent one, the choice of with new beta function terms to include in a more advanced analysis seems to be all but a trivial one.

\end{document}